\documentclass[prx,twocolumn,superscriptaddress]{revtex4}

\usepackage{amsmath,amssymb,amsfonts}
\usepackage{bm}
\usepackage{graphicx}

\begin{document}

\title{Emergent non-Fermi liquid at the quantum critical point of a topological phase transition
in two dimensions
}

\author{Hiroki \surname{Isobe}}
\affiliation{Department of Applied Physics, University of Tokyo,
Bunkyo, Tokyo 113-8656, Japan}

\author{Bohm-Jung \surname{Yang}}
\affiliation{RIKEN Center for Emergence Matter Science (CEMS),
Wako, Saitama 351-0198, Japan}

\author{Andrey \surname{Chubukov}}
\affiliation{William I. Fine Theoretical Physics Institute and School of Physics and Astronomy,
University of Minnesota, Minneapolis, MN 55455, USA}

\author{J\"{o}rg \surname{Schmalian}}
\affiliation{Institutes for Theory of Condensed Matter and for Solid State Physics,
Karlsruhe Institute of Technology, D-76131 Karlsruhe, Germany}

\author{Naoto \surname{Nagaosa}}

\affiliation{Department of Applied Physics, University of Tokyo,
Bunkyo, Tokyo 113-8656, Japan}

\affiliation{RIKEN Center for Emergence Matter Science (CEMS),
Wako, Saitama 351-0198, Japan}


\begin{abstract}
We study the effects of Coulomb interaction between
2D Weyl fermions with anisotropic dispersion which displays relativistic dynamics along one direction and Newtonian dynamics along the other.
 Such a dispersion can be realized
 in phosphorene under electric field or strain, in TiO$_2$/VO$_2$ superlattices, and, more generally,
at the quantum critical point between a nodal semimetal and an insulator in systems with a chiral symmetry.
Using the one-loop renormalization group approach in combination with the large-$N$ expansion,
we find that the system displays interaction-driven non-Fermi liquid behavior
in a wide range of intermediate frequencies and
 marginal Fermi liquid behavior at the smallest frequencies.
In the non-Fermi liquid regime, the
quasiparticle residue $Z$ at energy $E$ scales as $Z \propto E^a$ with
$a >0$, and the parameters of the fermionic dispersion  acquire
anomalous dimensions. In the marginal Fermi-liquid regime, $Z \propto (|\log E|)^{-b}$ with universal $b = 3/2$.
\end{abstract}


\maketitle
\textit{Introduction.---}%
After the discovery of time-reversal invariant topological band insulators~\cite{hasan,qi},
the notion of topological states of matters has been extended
to a broad class of systems.
In particular, recent  studies
of three-dimensional (3D) Weyl and Dirac semimetals with nodal points have demonstrated
that these systems
 also possess
 quantized topological invariants (or topological charges) and  associated topological surface states~\cite{wan2011,Matsuura,Zhao1,Zhao2,Chiu,Morimoto_Z2,yang_DiracSM2014,yang_DiracSM2015,burkov}.
Since the topological invariant assigned to each nodal point
guarantees its stability, the transition from a topological semimetal
to an insulator can be achieved
when pairs of nodal
points with
 opposite topological charges merge at the same momentum~\cite{Murakami}.
 Hence the quantum critical point (QCP) of semimetal-insulator transitions
should have emerging gapless degrees of freedom with zero topological charge~\cite{yang2013,yang2014}.

Because the topological charge of a nodal point
is solely determined by the energy dispersion around it, the vanishing
 of a topological charge at a QCP implies that at this point the dispersion of low energy excitations must become
 unconventional. Indeed, it has  recently been shown that
a new type of fermionic excitations, dubbed a 3D anisotropic Weyl fermion (AWF),
appears at the QCP between
a 3D Weyl semimetal and an insulator~\cite{Murakami,yang2013,yang2014}.
 A 3D AWF has an anisotropic energy dispersion,
which is quadratic in one direction
 and linear in the other two orthogonal directions.
 Such a dispersion  brings about
highly unusual quantum critical behavior.
Most notably,  quantum fluctuations of 3D AWFs
 screen the long-range Coulomb interaction and make it anisotropic, however
 the long-ranged nature of the interaction is preserved.
The screened anisotropic Coulomb potential
becomes an irrelevant perturbation in the low-energy limit, i.e.,
low-energy fermions remain free quasiparticles~\cite{yang2014}.

In this letter, we describe a semimetal-insulator transition and
 associated quantum criticality in a system of  two-dimensional (2D) AWF with long-range Coulomb interaction.
The role of the Coulomb interaction in 2D nodal semimetals  $V(\bm{q})\propto1/|\bm{q}|$
has been widely studied with particular emphasis on graphene~\cite{Graphene,aleiner,Son,Gonzalez}.
The conclusion is that at strong coupling, the Coulomb interaction
 generates anomalous exponents~\cite{Son}. This behavior, however, holds only at intermediate frequencies because the
  dimensionless coupling flows towards smaller values, and below a certain energy the system necessary
  enters into a weak coupling regime. In
   this regime the renormalizations are only logarithmical (marginal)~\cite{Gonzalez,Sheehy,Herbut,goswami2011,hosur2012,isobe2012,isobe2013}, and the quasiparticle residue tends to a finite value at zero energy, i.e., at  smallest frequencies graphene preserves Fermi-liquid behavior.
 As a consequence, interactions dress physical
  observables, like the optical conductivity, only
   by extra logarithmic
   factors~\cite{Sheehy,Herbut,Mishchenko,Sheehy09}.

We argue that
 the behavior  changes
 fundamentally when a semi-metal is brought to the quantum critical point of the semimetal-insulator transition,
 at which pairs of nodal points merge.
In this case, the low energy excitations around
a gapless point are 2D AWF with linear dispersion in one direction and quadratic in the other~\cite{montambaux,montambaux2,dietl2008,ezawa,pardo2009,pardo2010,banerjee2009,delplace2010,banerjee2012}, i.e.,
 a 2D AFW displays simultaneously relativistic and Newtonian dynamics.

\begin{figure}
\centering
\includegraphics[width=\hsize]{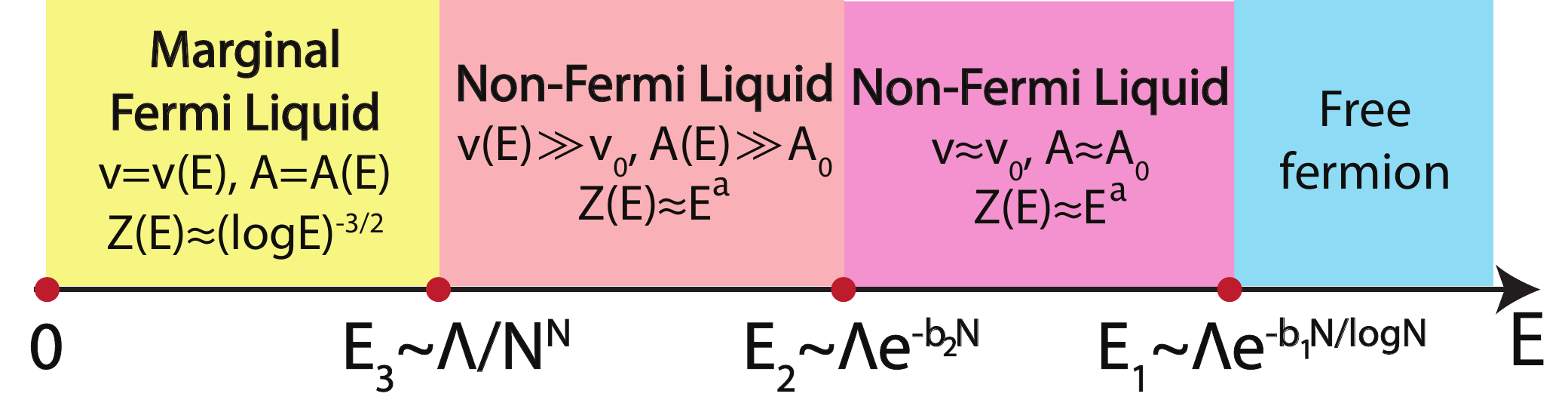}
\caption{(Color online)
Evolution of quasiparticle properties as the energy scale varies.
There are three energy scales $E_{1}\sim \Lambda e^{-b_{1}N/\log N}$,
$E_{2}\sim \Lambda e^{-b_{2}N}$, and $E_{3}\sim \Lambda/N^{N}$ characterizing each region.
Here $b_{1,2}$ are constants of O(1), and $v(E)$, $A(E)$, $Z(E)$ are the velocity, the inverse mass,
and the quasiparticle residue, which are running
as $E$ decreases. $\Lambda$ is of the order of the band width.
}
\label{fig:energyscale}
\end{figure}

We  analyze the effects of Coulomb interaction in 2D AWF's
 by combining a renormalization group (RG) analysis and a large $N$ expansion.
  We find (see Fig.~\ref{fig:energyscale}) that over a wide range of energies the screened Coulomb interaction is a relevant perturbation, and the system is in the strong coupling limit and displays non-Fermi liquid behavior with power-law energy dependence of the quasiparticle residue $Z \propto E^{a}$, where $a = O((\log{N})/N)$.  This behavior starts at
  $E_1 \sim \Lambda e^{-b_{1}N/\log N}$, where $\Lambda$ is of the order of the bandwidth and $b_{1} = O(1)$, and extends down to very low energy $E_3 \sim \Lambda/ (N^N)$.
In the subrange $E_3 < E < E_2 < E_1$, where $E_2 \sim \Lambda e^{-b_{2}N}, ~b_{2} = O(1)$,  the parameters of the fermionic dispersion (the velocity and the effective mass) also become energy dependent and vary as powers of $E$ with anomalous exponents $O(1/N)$.
At even smaller energies $E < E_3$, the system crosses over to weak coupling behavior.  However, contrary to
 the case of 2D nodal semimetals, 2D AWF do not become free particles at the smallest frequencies. Instead, in the limit $E \to 0$, the system displays marginal Fermi liquid behavior with universal, $N$-independent quasiparticle weight $Z (E) \propto (\log E)^{-3/2}$.

\textit{The model.---}%
Non-interacting 2D AWF are described by
\begin{equation}
H_0  = -A \partial_{x}^2 \tau_x -iv \partial_{y} \tau_y,
\end{equation}
where $1/(2A) >0$ is the mass along $x$ direction and $v$ is the velocity along $y$ direction.
The Pauli matrices $\tau_{x, y}$ are used to denote the valence and conduction bands.
The absence of $\tau_{z}$ in the Hamiltonian ensures the chiral symmetry of the system, thus a nodal point can have an integer winding number.
The semimetal-insulator transition can be described by
adding to $H_0$ a perturbation term $m\tau_{x}$ with a constant $m$.
Depending on the sign of $m$, the system becomes either a gapped insulator $(m>0)$
or a semimetal $(m<0)$ with two nodal points on the $k_{x}$ axis with the winding numbers $\pm1$, respectively.
The distance between two nodal points decreases
as $|m|$ is reduced, and at $m=0$
two nodal points merge, resulting in a gapless point
with a zero topological charge~\cite{montambaux,montambaux2}.
The anisotropic dispersion leads to the density of states $\rho_{\text{QCP}}(E)\propto \sqrt{E}$,
which is obviously enhanced in the low energy limit as compared to
the case of a semimetal for which $\rho_{\text{SM}}(E)\propto E$.
This suggests that interaction effects are  most relevant at the QCP.

We consider long-range Coulomb interaction between 2D AWF. The corresponding
 effective action is
\begin{align}
S = &\int d\tau d^2 x \psi_a^\dagger [(\partial_\tau + ig\phi ) +H_{0}] \psi_a
+ \frac{1}{2}\int d\tau d^3 x (\partial_{i} \phi)^{2} ,
\end{align}
where
$\psi_{a}$ describes a two-component fermion field with the subscript $a=1,...,N$
labeling the species of AWF, and
$\phi$ is a bosonic field which one obtains via Hubbard-Stratonovich transformation of the instantaneous Coulomb potential.
 The subscript $i=x,y,z$ and the summation over repeated indices is implied.
Observe that the bosonic field $\phi$ is defined in 3D space whereas the
electron is confined to a 2D plane.
Once the
 $z$-dependence of $\phi$,
 is integrated out, the Coulomb potential in momentum space becomes
$V(\bm{q})\propto1/|\bm{q}|$.
The dimensionful
boson-fermion coupling associated with Coulomb potential is $g = e/\sqrt{\varepsilon}$, where $e$ is the electric charge and $\varepsilon$ is determined by the dielectric constant.
 The corresponding dimensionless coupling $\alpha$, which appears in perturbation theory,  is the ratio of the Coulomb potential $E_c \sim A^{-1} v g^2$ and the electron kinetic energy
$E_{kin} \sim A^{-1} v^2$:  $\alpha \equiv \frac{E_{c}}{E_{\text{kin}}}= \frac{g^2}{v}$.
 To control the theory analytically, we extend the model to $N$ fermionic flavors and consider the large $N$ limit.
 At large $N$, the dimensionless coupling constant becomes
  $\alpha_N = N \alpha$.

\begin{figure}
\centering
\includegraphics[width=\hsize]{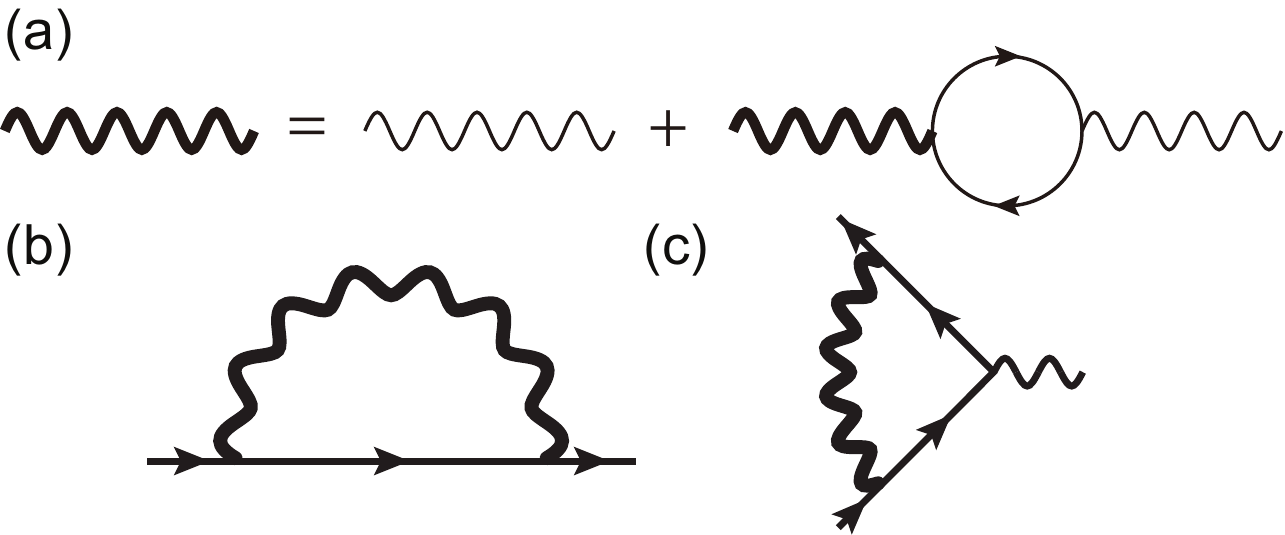}
\caption{
(a) Feynman diagrams representing the RPA boson propagator (bold wavy line).
Each fermion loop, accompanied by a factor $N$, is resummed in the boson propagator.
(b) electron self-energy and (c) the vertex function.
}
\label{fig:diagrams}
\end{figure}

\textit{Bosonic and fermionic propagators.---}%
We follow the same strategy as in earlier approaches on large $N$ theories of quantum-critical behavior of itinerant fermions~\cite{Abanov,Son}
   and compute fermionic and bosonic self-eneries in a self-constent scheme. Namely, we first compute the {\it one-loop} bosonic self-energy (bosonic polarization operator $\Pi_1 (\Omega,\bm{q})$) as the latter contains $N$ (and also, as we will see, contains smaller
    power of $q_y$ compared to the bare term $D_{0}^{-1}=(q^2_x + q^2_y)^{1/2}$), then use the dressed dynamical bosonic propagator
    \begin{align}
D^{-1}_1 (\Omega,\bm{q}) = D_{0}^{-1}(\bm{q})-N \Pi_1 (\Omega,\bm{q}),
\end{align}
    to compute the fermionic self-energy  and corrections to the one-loop $\Pi_1 (q, \Omega)$ within the
    $1/N$ expansion.  We show that the corrections depend logarithmically on the running energy and solve the RG equations for the full propagators using $1/N$ as a control parameter.

The evaluation of the one-loop polarization bubble is rather involved. We present the details in the Supplemental Material (SM) and here
 list the result. We find that
  $\Pi_1 (\Omega,\bm{q})$ can be expressed as
\begin{align}
\label{eq:ansatz}
\Pi_1 (\Omega,\bm{q}) =
- \alpha \left[
\frac{d_{x}A^{1/2}q_{x}^{2}}{\Delta (\Omega, \bm{q})^{1/4}}
+\frac{d_{y} A^{-1/2} v^{2} q_{y}^{2}}{\Delta (\Omega, \bm{q})^{3/4}}
\right],
\end{align}
where $\Delta (\Omega, \bm{q}) = \Omega^{2} + cA^2 q_x^4 + v^2 q_y^2$, and $d_x$, $d_y$, and $c$ are constants,
whose explicit values we present in SM.
This $\Pi_1(\Omega,\bm{q})$ matches the exact results, which we obtained analytically, in the three limits:
 (i) $\Pi_1(\Omega=q_{y}=0)
= -\frac{g^{2}}{16v}|q_{x}|$,  (ii) $\Pi_{1}(|\Omega| \gg Aq_x^2,q_{y}=0)
 = - \frac{\alpha}{8\sqrt{\pi}} \frac{\Gamma(3/4)}{\Gamma(9/4)} \frac{\sqrt{A} q_{x}^{2}}{\sqrt{\Omega}}$, (iii) $\Pi_{1}(q_{x}=0)
= - \frac{\alpha}{6\sqrt{\pi}} \frac{\Gamma(5/4)}{\Gamma(3/4)} \frac{v^{2}q_{y}^{2}/\sqrt{A}}{(\Omega^{2}+v^{2}q_{y}^{2})^{3/4}}$.
Observe that  while  $\Pi_1(\Omega=q_{y}=0) \propto |q_x|$ has the same functional dependence as  $D_{0}^{-1}(q_x) = |q_x|$,
$\Pi_1 (q_x =0, \Omega \sim v q_y) \propto |q_y|^{1/2}$ is parametrically larger than $D_{0}^{-1}(q_y) = |q_y|$ at small $q_y$ (Ref.~\cite{extra}).

We now use the bosonic propagator with $\Pi_1 (\Omega, {\bf q})$ included and compute the one-loop fermionic self-energy $\Sigma_1(\omega,\bm{k})$
 and vertex correction $\delta g_1$. The corresponding diagrams are shown in Fig.~\ref{fig:diagrams}.
  For the self-energy, we obtain at large $\alpha_N = N g^2/v$ (see SM for details.)
\begin{align}\label{eqn:electronselfenergy}
\Sigma_1(\omega,\bm{k})
&= (-ig)^{2}\int\frac{d\Omega d^{2}q}{(2\pi)^{3}}
G_{0}(\omega+\Omega,\bm{k}+\bm{q})D(\Omega,\bm{q}) \notag \\
& \equiv \Sigma_\omega \cdot i\omega - \Sigma_{k_x} \cdot Ak_x^2 \tau_x - \Sigma_{k_y} \cdot v k_y \tau_y ,
\end{align}
in which
\begin{align}
\Sigma_\omega = \gamma_z l,
~ \Sigma_{k_x} = \Sigma_\omega +  \gamma_A l,~ \Sigma_{k_y} = \Sigma_\omega +  \gamma_v l,
\label{ac_2}
\end{align}
where $l = \log(\Lambda/E)$, $\Lambda$ is the upper energy cutoff, and $E$ is the largest of $(|\omega|, v_F |k_y|, A k^2_x)$.
The parameters are $\gamma_z = \frac{\sqrt{15}}{\pi^{3/2}}\log {\alpha_N}/N$,  $\gamma_A = 0.1261/N$, and $\gamma_v = 0.3625/N$.
We see that the quasiparticle residue $Z = (1 + \partial \Sigma/\partial i\omega)^{-1}$ and the parameters of electronic dispersion acquire
logarithmically singular $1/N$ corrections. Observe that the correction to $Z$ is stronger than the corrections to $v$ and $A$ by $\log {\alpha_N}$.

For the vertex correction we find at vanishing external momentum and frequency
\begin{align}
\delta g _1 &= -g^2 \int \frac{d^3k}{(2\pi)^3} D(k) G_0 (k) G_0 (k) \equiv \Sigma_\omega.
\end{align}
This is consistent with the Ward-Takahashi identity. The calculation of $\delta g_1$ generally requires care as $\delta g_1$ has terms which depend on how the limit of zero momentum and frequency is taken~\cite{landau}.  However these terms are non-logarithmical and can be safely neglected at large $N$.

\textit{Renormalization group analysis.---}%
One can verify that higher-order corrections contain higher powers of $l$. To sum up the series of logarithms, we can either use the Wilsonian shell RG analysis~\cite{shell_RG}, or express  the full self-energy $\Sigma (\omega, {\bf k})$, the full vertex $g$, and  the full polarization bubble self-consistently, via
 full Green's functions and full vertices, and then represent $\Sigma (l)$, $g(l)$, and $\Pi(l)$ as  integrals $\int^l d l'$ over running $l'$, and obtain RG equations by taking derivatives with respect to the upper limit~\cite{rg}.  Either way we obtain the same set of RG equations for running $Z$,
 $v$, and $A$. We also verified that the product $g Z$ is not renormalized, as it is required by the condition that the electric charge is a
  conserved quantity.  The RG equations for $Z(l)$, $v(l)$, and $A(l)$ are (${\dot X} = dX/dl$)
\begin{align}
{\dot Z} (l) = -\gamma_z (l) Z (l),~{\dot v}(l) = \gamma_v v (l), ~{\dot A} (l) = \gamma_A A(l)
\label{ch_3}
\end{align}
 where $\gamma_z (l) = \gamma_z + \frac{\sqrt{15}}{\pi^{3/2}}\frac{1}{N} \log{\frac{v}{v(l)}}$.
Solving these equations and using $l = \log ({\Lambda/E})$   we obtain $v(E)$, $A(E)$, and $Z(E)$ at energy $E$ in the form
\begin{align}
\frac{v (E)}{v} = \left(\frac{\Lambda}{E}\right)^{\gamma_v}, \frac{A(E)}{A} = \left(\frac{\Lambda}{E}\right)^{\gamma_A},
Z(E) = \left(\frac{\Lambda}{E}\right)^{-{\gamma_z} + \frac{\sqrt{15}}{\pi^{3/2}}\frac{\gamma_v}{N} l}.
\label{ac_1}
\end{align}
The analysis of Eq.~(\ref{ac_1}) shows that there are three energy scales characterizing the system's behavior.
 At high energies, $E > E_1 = \Lambda e^{-b_{1}N/\log N}$, where $b_{1} = O(1)$,
 the dependence of the fermionic propagator on $E$ is weak, i.e., fermions behave as almost
  free quasiparticles.   At $E_2 < E < E_1$, where $E_2 = \Lambda e^{-b_{2}N}, ~b_{2} = O(1)$, $v$ and $A$ remain close to their bare values, but the
   quasiparticle residue becomes strongly $E$-dependent, and the fermionic
   propagator at the typical energy $E$ acquires a non-Fermi-liquid form with anomalous exponent $\gamma_z$, i.e.,
   $G \propto 1/E^{1 -\gamma_z}$.  This behavior holds also at energies
    below $E_2$, but now $v$ and $A$ grow as powers of $\Lambda/E$ with anomalous exponents $\gamma_v$ and $\gamma_A$, respectively.
The presence of anomalous dimensions in the theory
implies that physical observables, such as the
 specific heat, the compressibility, the diamagnetic susceptibility, etc.,  show
unusual temperature dependencies, as shown below.

The strong coupling results differ quantitatively but not qualitatively from the case of graphene.  In both cases,  Coulomb interaction gives rise to anomalous exponents for the quasiparticle residue and the fermionic dispersion.

\textit{Weak coupling limit.---}The behavior described by Eq.~(\ref{ac_1}) holds as long as the dressed dimensionless coupling $\alpha_N = (gZ)^2 N/v$
 remains large. The bare value of $\alpha_N$ is of order $N$, however $v(E)$ grows upon the flow towards lower energies,
 and eventually, at $E < E_3 = \Lambda/ N^N$, the dimensionless coupling $\alpha_N$ becomes smaller than one. Once this happens, the RG equations have to be modified because, e.g., bare $|q_x|$ in the bosonic propagator becomes larger than the $|q_x|$ term in the polarization operator.   We evaluated the one-loop self-energy and vertex corrections at small $\alpha_N$ and found that they are again logarithmical,
 but the RG parameter $l$ is now  $l = \log (E_{m}/E)$, where $E_m = g^2 N^2/A$ is the energy scale below which
 the $\sqrt{|q_{y}|}$ term in the polarization dominates over the bare term $|q_{y}|$ in the boson propagator. The
    prefactors $\gamma_z, \gamma_v$ and $\gamma_A$ are different from those in Eq. (\ref{ac_2}) and are given by
  \begin{align}
  \gamma_z = \frac{3 \alpha_N}{8\pi^2 N},~\gamma_v = \frac{\alpha_N}{4\pi^2 N},~\gamma_A = \frac{\alpha_N |\log{\alpha_N}|}{2\pi^2 N},
 \label{ch_4}
  \end{align}
 where $\alpha_N = g^2 N/v$ and $g Z$ is not renormalized.  Higher-order corrections give higher powers of logarithms, and performing the same analysis as in the strong coupling limit, we find that 
  $Z (l)$, $v(l)$, and $A(l)$ satisfy the same RG equations (\ref{ch_3}) as in the strong-coupling limit, but with $\gamma$'s from Eq.~(\ref{ch_4}).
 Solving these equations we obtain at smallest energies (largest $l$)
  \begin{align}
  v(l) =  \frac{g^2}{4\pi^2} l, ~ Z(l) = l^{-3/2},~ A(l) = A e^{\log^2 l}.
 \label{ch_5}
  \end{align}
We see that the quasiparticle $Z$ does not reduce to a constant in the limit $\alpha_N \to 0$ but keeps decreasing, even at the smallest $E$.
 At the same time, we see that the series of logarithms at weak coupling do not generate anomalous dimensions, i.e., $v (l)$ and $Z(l)$ behave as powers of $l$ (the inverse mass $A$ has a somewhat more complex dependence on $l$, but still there is no anomalous dimension.)
  The logarithmic form of $Z(l)$ implies that the fermionic Green's function at the smallest energies behaves as $G(E) \sim (\log E)^{-3/2} /E$. Such behavior is generally termed as
  {\it a marginal Fermi liquid}.

This last result shows that the weak coupling behavior of AWF is {\it qualitatively} different from that in graphene.
In graphene, the quasiparticle $Z$ factor tends to a finite value at zero energy~\cite{Son},
i.e. the system retains Fermi-liquid behavior with well defined quasiparticles.
AWF, on the contrary, do not become sharp quasiparticles, even at the lowest energies.

\begin{figure}
\centering
\includegraphics[width=0.8\hsize]{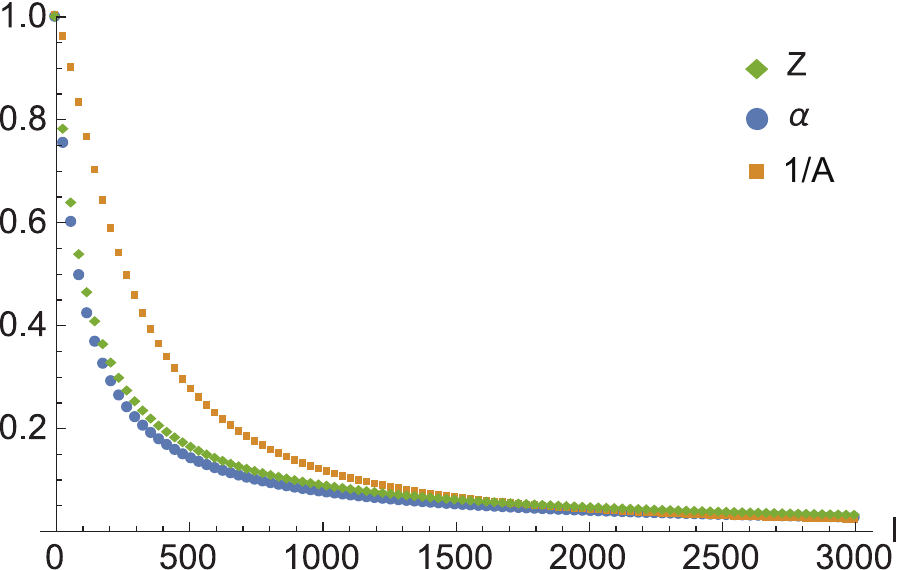}
\caption{(Color online)
RG flow of the coupling constant
$\alpha \propto 1/v $, the effective mass $1/A$, 
and the quasiparticle residue $Z$
 for 2D AWF.
We set the initial values $\alpha_N = 4$ and $A = 1$ and the number of flavors $N=4$.
$\alpha$, $1/A$, $Z$ all flow to zero.
The  quasiparticle residue has power-law dependence $Z (E) \propto E^a$ at intermediate energies, 
typical for a non-Fermi liquid, and scales as $Z(E) \propto (1/|\log E|)^{3/2}$  at the lowest energies, 
i.e., at vanishing $E$ the system displays 
marginal Fermi liquid behavior.
}
\label{fig:rgflow}
\end{figure}

To verify our analytical analysis, we obtained the RG flow of $Z$, $\alpha_N \propto 1/v$, and $1/A$ numerically for $N=4$. (See
 Fig.~\ref{fig:rgflow}.)
 We clearly see that $v$ and $A$ increase upon the system flows to lower energies (higher $l$), whereas the
quasiparticle residue $Z$ decreases, initially by a power law, and then nearly flattens at the largest $l$.

\textit{Physical observables.-}
Now we obtain scaling relations for physical observables.
In general, an operator $O$ with the scaling dimension $z_{O}$ obeys
\begin{equation}
O\left(\mathbf{k},\omega,v,A\right)=z_{O}O\left(\mathbf{k}\left(l\right),\omega\left(l\right),v\left(l\right),A\left(l\right)\right),
\end{equation}
where
 $\mathbf{k}\left(l\right)=(z_{x}\left(l\right)k_{x},e^{l}k_{y})$
and $\omega\left(l\right)=z_{\omega}\left(l\right)\omega$ are the
running momenta and frequencies. $z_{x,\omega}$ are determined by
$\frac{d\log z_{\omega}}{dl}=1-\gamma_{A}$ and $\frac{d\log z_{x}^{2}}{dl}=1+\gamma_{A}-\gamma_{v}$.
In the case of the particle density $O=n$, $z_{O=n}=e^{-l}z_{x}^{-1}$.
Standard scaling arguments then yield, at strong coupling, for the compressibility $\kappa=\frac{\partial n}{\partial\mu}\propto T^{1/2+\phi}$
with $\phi=\gamma_{v}+\frac{1}{2}\gamma_{A}\approx0.4255/N$. Similarly,
the heat capacity
 $C\propto T^{\frac{3}{2}+\phi}$, and
the diamagnetic susceptibility
$\chi_{{\rm dia}}\propto-T^{-\frac{1}{2}-\phi}$~\cite{Sheehy}. Particularly interesting
behavior is expected
 for the anisotropic transport and optical properties.
The real part $\sigma_{\alpha}\left(\omega\right)$ of the optical
conductivity for non-interacting AWF is anisotropic with $\sigma_{x,y}^{0}\left(\omega\right) \propto \left(\omega / \omega_{0}\right)^{\pm1/2}$,
where $\omega_{0}=v^{2}/A$
and the upper (lower) sign stands for the $x$ ($y$) direction. Using
gauge invariance, one finds that the scaling dimensions of the optical
conductivity are different for the two directions: $z_{\sigma_{x}}=z_{x}e^{-l}$
and $z_{\sigma_{y}}=z_{\sigma_{x}}^{-1}$. In the strong coupling
regime, we immediately obtain that
\begin{equation}
\sigma_{x,y}\left(\omega\right)\propto N \frac{e^2}{\hbar} \left(\frac{\omega}{\omega_0}\right)^{\pm\left(\frac{1}{2}+\phi_{\sigma}\right)}
\end{equation}
with $\phi_{\sigma}=\gamma_{v}-\gamma_{A}/2\approx0.299/N$. Thus,
the anisotropy of the optical conductivity is amplified by
 strong interactions. In the weak coupling limit holds instead:
\begin{equation}
\sigma_{x,y}\left(\omega\right)\propto N\frac{e^{2}}{\hbar}\left(\frac{\omega e^{\log^{2}\log\frac{\Lambda}{\omega}}}{\omega_{0}\log^{2}\frac{\Lambda}{\omega}}\right)^{\pm\frac{1}{2}}.
\end{equation}

\textit{Conclusion.---}%
In this paper  we studied quantum critical behavior
of 2D interacting AWF. We showed that the interplay between the anisotropic
electron dispersion and the long-range Coulomb interaction
generates a highly anisotropic structure of the screened Coulomb potential,
which, in turn, induces unconventional behavior of electrons.
 We demonstrated that interacting 2D AWFs display non-Fermi liquid at intermediate energies, with various anomalous physical properties, and marginal Fermi liquid behavior at the smallest energies.

There are several candidate materials for 2D AWF.
In deformed graphene~\cite{montambaux,montambaux2,dietl2008},
pressured organic conductor $\alpha$-(BEDT-TTF)$_{2}$I$_{3}$~\cite{organic}, 
and artificial lattices of cold atoms~\cite{coldatom}, 
2D AWF emerge via the merging of two Dirac points.
In TiO$_2$/VO$_2$ nanostructures 2D AWF were predicted to be intrinsic low energy excitations
due to the peculiar symmetry of the system~\cite{pardo2009,pardo2010,banerjee2009,delplace2010,banerjee2012}.
2D AWF  were also predicted theoretically to exist under external electric field or strain 
in black phosphorous (a system with a few layers of phosphorene) ~\cite{ezawa,phosphorene}.
This prediction has been confirmed in a recent angle-resolved photo-emission study~\cite{kskim}.
We believe that intriguing non-Fermi liquid physics can be probed in
these systems through a systematic investigation of quasiparticle properties.

B.-J. Y and N. N. were supported by a Grant-in-Aid for Scientific Research on
Innovative Areas ``Topological Materials Science'' (KAKENHI Grant No. 15H05853)
and Kiban S “Emergent electromagnetism in magnets” (KAKENHI Grant No. 24224009).
H. I. was supported by a Grant-in-Aid for JSPS Fellows. A. C was supported by the DOE grant \# CON000000051512, and
J.S. is supported by  the Deutsche Forschungsgemeinschaft through grant SCHM 1031/4-1.

\textit{Note added.---}After the completion of our paper, we became aware of the independent work by Gil Young Cho and Eun-Gook Moon on the same problem.

\onecolumngrid
\noindent
\hrulefill

\section*{Supplemental Material}

The effective action describing the two-dimensional anisotropic Weyl semimetals in momentum space with Matsubara frequency is 
\begin{equation}
S=S_{F}+S_{B}+S_{g},
\end{equation}
where
\begin{gather}
S_{F}=\int \frac{d\omega}{2\pi}\frac{d^{2}k}{(2\pi)^{2}} \psi^{\dagger}(\omega,\bm{k}) (-i\omega+Ak_{x}^{2}\tau_{x}+v k_{y}\tau_{y}) \psi(\omega,\bm{k}), \\
\label{eqn:boson}
S_{B}=\frac{1}{2}\int \frac{d\omega}{2\pi}\frac{d^{3}k}{(2\pi)^{2}} \phi^{\dagger}(\omega,\bm{k}) ( k_x^2 +  k_y^2 + k_z^2) \phi(\omega,\bm{k}), \\
S_{g}=\int \frac{d\omega}{2\pi}\frac{d\omega'}{2\pi}\frac{d^{2}k}{(2\pi)^{2}}\frac{d^{2}k'}{(2\pi)^{2}}
\psi^{\dagger}(\omega,\bm{k})[ig\phi(\omega-\omega',\bm{k}-\bm{k}')]\psi(\omega',\bm{k}').
\end{gather}
The fermion propagator $G_0(\omega, \bm{k})$ and the boson propagator reduced to the 2D plane $D_0(\bm{q})$ are given by
\begin{gather}
G_{0}(\omega,\bm{k}) = \frac{1}{-i\omega+Ak_{x}^{2}\tau_{x}+v k_{y}\tau_{y}} , \\
D_{0}(\bm{q}) = \int \frac{dk_z}{2\pi} \frac{1}{ k_x^2 +  k_y^2 + k_z^2}
= \frac{1}{2\sqrt{q_{x}^{2}+ q_{y}^{2}}}. 
\end{gather}
The vertex of fermion and boson propagators gives $(-ig)$.

\section{\label{sec:boson} Polarization function}

The polarization at one-loop level is given by
\begin{align}
\Pi(\Omega,\bm{q}) & =(-1)(-ig)^{2}\int\frac{d\omega}{2\pi}\int \frac{d^{2}k}{(2\pi)^{2}}\text{Tr}[G_{0}(\Omega+\omega,\bm{q}+\bm{k})G_{0}(\omega,\bm{k})] \notag \\
& = 2g^{2}\int\frac{d\omega}{2\pi}\int \frac{d^{2}k}{(2\pi)^{2}}
\frac{-\omega(\Omega+\omega)+A^{2}k_{x}^{2}(k_{x}+q_{x})^{2}+v^{2}k_{y}(k_{y}+q_{y})}{[(\Omega+\omega)^{2}+A^{2}(q_{x}+k_{x})^{4}+v^{2}(q_{y}+k_{y})^{2}][\omega^{2}+A^{2}k_{x}^{4}+v^{2}k_{y}^{2}]}.
\end{align}
By using Feynman parameterization
\begin{equation}
\frac{1}{AB}=\int_{0}^{1}dx\frac{1}{[xA+(1-x)B]^{2}},
\end{equation}
we can perform the $\omega$ and $k_y$ integrations analytically  to obtain
\begin{align}
\Pi \left(\Omega,\frac{q_{x}}{\sqrt{A}},\frac{q_{y}}{v} \right)
&= \frac{g^{2}}{4\pi^{2}v\sqrt{A}}\int_{-\infty}^\infty dk_{x}\int_{0}^{1}dx
\frac{-2x(1-x)q_{y}^{2} + k_{x}^{2} (k_{x}+q_{x})^{2} - x (k_{x}+ q_{x})^{4} - (1-x) k_{x}^{4}}
{x(1-x)\Omega^{2} + x(1-x) q_{y}^{2} + x (k_{x}+q_{x})^{4} + (1-x) k_{x}^{4}}.
\end{align}

\subsection{$\Omega = 0$ and $q_y = 0$}

Assuming $\Omega=q_{y}=0$ we have
\begin{align}
\Pi \left(\Omega = 0,\frac{q_{x}}{\sqrt{A}},\frac{q_{y}}{v} =\log 0 \right)
&= \frac{g^{2}}{4\pi^{2}v\sqrt{A}}\int_{-\infty}^\infty dk_{x}\int_{0}^{1}dx
\frac{k_{x}^{2} (k_{x}+q_{x})^{2} - x (k_{x}+ q_{x})^{4} - (1-x) k_{x}^{4}}
{x (k_{x}+q_{x})^{4} + (1-x) k_{x}^{4}} \notag \\
&= \frac{g^{2}}{4\pi^{2}v\sqrt{A}}\int_{-\infty}^\infty dk_{x}
\left\{-1+\frac{k_{x}^{2}(k_{x}+q_{x})^{2}}
{[(k_{x}+q_{x})^{4}-k_{x}^{4}]}\log\left[\frac{(k_{x}+q_{x})^{4}}{k_{x}^{4}}\right]\right\}. 
\end{align}
Using the relation
\begin{equation}
\int_{-\infty}^{\infty} dx \left\{1-\frac{x^{2}(x+1)^{2}}
{[(x+1)^{4}-x^{4}]}\log\left[\frac{(x+1)^{4}}{x^{4}}\right]\right\} = \frac{\pi^{2}}{4}, 
\end{equation}
we obtain
\begin{equation}
\Pi(\Omega=0,q_{x},q_{y}=0) = -\frac{\alpha}{16}|q_{x}|.
\end{equation}

\subsection{$\Omega \gg Aq_x^2$ and $q_y=0$}

For $\Omega \gg Aq_x^2$ and $q_y=0$, we consider the lowest order contribution in $(Aq_x^2/\Omega)$, and then the polarization becomes
\begin{align}
\Pi \left(\Omega, \frac{q_{x}}{\sqrt{A}}, \frac{q_{y}}{v}=0 \right)
&= \frac{g^{2}}{4\pi^{2}v\sqrt{A}}\int_{-\infty}^\infty dk_{x}\int_{0}^{1}dx
\frac{k_{x}^{2} (k_{x}+q_{x})^{2} - x (k_{x}+ q_{x})^{4} - (1-x) k_{x}^{4}}
{x(1-x)\Omega^{2} + x (k_{x}+q_{x})^{4} + (1-x) k_{x}^{4}} \notag \\
&= \frac{g^{2}}{4\sqrt{2}\pi v} \frac{q_x^2}{\sqrt{A\Omega}} \int_{0}^{1}dx
\frac{1-12x(1-x)}{[x(1-x)]^{1/4}}   \notag \\
&= \frac{g^{2}}{4\sqrt{2}\pi v} \frac{q_x^2}{\sqrt{A\Omega}}
\left[ B\left( \frac{3}{4}, \frac{3}{4} \right) -12 B\left( \frac{7}{4}, \frac{7}{4} \right) \right].
\end{align}
Here $B(a,b)$ is the beta function 
and therefore we obtain
\begin{equation}
\Pi(\Omega, q_x, q_y = 0) = - \frac{\alpha}{8\sqrt{\pi}} \frac{\Gamma(3/4)}{\Gamma(9/4)} \frac{\sqrt{A} q_x^2}{\sqrt{\Omega}} \left[ 1 + O\left(\frac{A q_x^2}{\Omega}\right) \right].
\end{equation}

\subsection{$q_{x}=0$}

For $q_x=0$, we can calculate the polarization function including the frequency dependence: 
\begin{align}
\Pi \left(\Omega, \frac{q_{x}}{\sqrt{A}} = 0, \frac{q_{y}}{v} \right)
&= \frac{g^{2}}{4\pi^{2}v\sqrt{A}}\int dk_{x}\int_{0}^{1}dx
\frac{-2x(1-x)q_{y}^{2}}
{[x(1-x)(\Omega^{2}+q_{y}^{2})+k_{x}^{4}]}
\notag \\
&= -\frac{g^{2}}{2\sqrt{2}\pi v \sqrt{A}}
\frac{q_{y}^{2}}{(\Omega^{2}+q_{y}^{2})^{3/4}}
\int_{0}^{1}dx x^{1/4} (1-x)^{1/4}, \notag \\
&= -\frac{g^{2}}{2\sqrt{2}\pi v \sqrt{A}}
\frac{q_{y}^{2}}{(\Omega^{2}+q_{y}^{2})^{3/4}}
B \left( \frac{5}{4}, \frac{5}{4} \right).
\end{align}
Therefore the polarization for $q_x = 0$ becomes
\begin{equation}
\Pi(\Omega, q_x = 0, q_y) = -\frac{\alpha}{6\sqrt{\pi}} \frac{\Gamma(5/4)}{\Gamma(3/4)} \frac{v^2 q_y^2 / \sqrt{A}}{(\Omega^2 + v^2 q_y^2)^{3/4}}.
\end{equation}

\subsection{\label{sec:ansatz} An ansatz for the polarization function}
We propose the following ansatz for the polarization function:
\begin{equation}
\label{eq:ansatz}
\tilde{\Pi}(\Omega,q_{x},q_{y})
= -\frac{d_x A^{1/2} q_{x}^{2}}{(\Omega^{2}+v^{2}q_{y}^{2}+cA^{2}q_{x}^{4})^{1/4}}
- \frac{d_y A^{-1/2} v^{2}q_{y}^{2}}{(\Omega^{2}+v^{2}q_{y}^{2}+cA^{2}q_{x}^{4})^{3/4}},
\end{equation}
where
\begin{equation}
d_{x}=\frac{1}{8\sqrt{\pi}}\frac{\Gamma(3/4)}{\Gamma(9/4)}, \quad
d_{y}=\frac{1}{8\sqrt{\pi}}\frac{\Gamma(5/4)}{\Gamma(7/4)}, \quad
c=\left(\frac{2}{\sqrt{\pi}}\frac{\Gamma(3/4)}{\Gamma(9/4)} \right)^{4}.
\end{equation}
This ansatz is consistent with the exact result of $\Pi$ in the following three limits; i.e., 
(i) $\Omega = 0$ and $q_y = 0$, 
(ii) $\Omega \gg Aq_x^2$ and $q_y=0$, and 
(iii) $q_x=0$. 

Since $\tilde{\Pi}(\Omega,q_{x}=0,q_{y})$ is consistent with the exact result,
let us check the behavior of $\tilde{\Pi}(\Omega,q_{x},q_{y}=0)$ more carefully
and compare it with the exact result:
\begin{align}
\tilde{\Pi}(\Omega,q_{x},q_{y}=0)
&\equiv \alpha |q_{x}| \ F_{1} \left(\frac{\Omega^{2}}{A^{2}q_{x}^{4}} \right),
\end{align}
where
\begin{align}
F_{1} (u)
= \frac{1}{8\sqrt{\pi}} \frac{\Gamma(3/4)}{\Gamma(9/4)}
\frac{1}{( u + c )^{1/4}}.
\end{align}
\begin{figure}
\centering
\includegraphics[width=8cm]{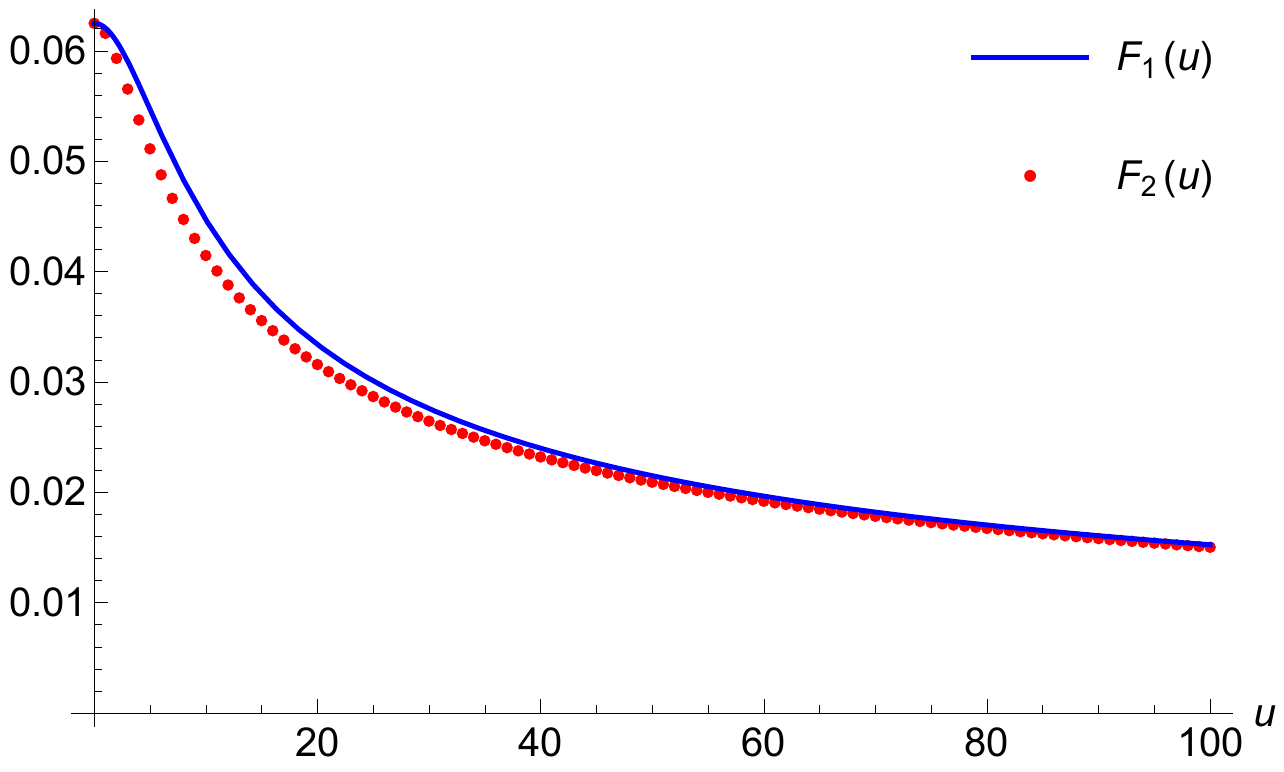}
\caption{
Comparison of the ansatz $F_{1}(u)$ (blue line) with the exact function $F_{2}(u)$ (red dots), which is numerically evaluated at each point.
The horizontal axis measures $u = 4\Omega/(Aq_{x}^{2})$.
}\label{fig:ansatz}
\end{figure}
On the other hand,  the exact formula of the polarization for $q_y = 0$ is 
\begin{equation}
\Pi(\Omega,q_{x},q_{y}=0) = -\alpha |q_{x}|F_{2} \left(\frac{4\Omega}{Aq_{x}^{2}} \right)
\end{equation}
where
\begin{equation}
F_{2}(u)=
-\frac{1}{8\pi^{2}}\int dzdp
\frac{(\sqrt{(z+1)^{4}+p^{2}}+\sqrt{(z-1)^{4}+p^{2}})[(z^2-1)^{2}+p^{2}-\sqrt{(z+1)^{4}+p^{2}}\sqrt{(z-1)^{4}+p^{2}}]}
{\sqrt{(z+1)^{4}+p^{2}}\sqrt{(z-1)^{4}+p^{2}}[u^{2}+(\sqrt{(z+1)^{4}+p^{2}}+\sqrt{(z-1)^{4}+p^{2}})^{2}]}. 
\end{equation}
Fig.~\ref{fig:ansatz} shows $F_{1}(u)$ (ansatz) and $F_{2}(u)$ (exact) as functions of $u= 4\Omega / (Aq_{x}^{2})$.
We confirm that the ansatz $F_{1}(u)$ is in good agreement with the numerically-estimated exact function $F_{2}(u)$.

\section{Self-energy at large $\alpha_N$}

By using the ansatz polarization function, the inverse boson propagator becomes
\begin{align}
D^{-1}(i\Omega,\bm{q})
&= D_{0}^{-1}(i\Omega,\bm{q})-N\tilde{\Pi}(i\Omega,\bm{q})
\notag \\
&= 2\sqrt{q_{x}^{2}+q_{y}^{2}}+
\alpha_N \left[
\frac{d_{x}A^{1/2}q_{x}^{2}}{(\Omega^{2}+v^{2}q_{y}^{2}+cA^{2}q_{x}^{4})^{1/4}}
+\frac{d_{y} A^{-1/2} v^{2} q_{y}^{2}}{(\Omega^{2}+v^{2}q_{y}^{2}+cA^{2}q_{x}^{4})^{3/4}}
\right], 
\end{align}
where $\alpha_N = N\alpha$, and the electron self-energy is
\begin{align}
\Sigma(i\omega,\bm{k})
&= (-ig)^{2} \frac{1}{\beta}\sum_{i\Omega}\int\frac{d^{2}q}{(2\pi)^{2}}
G_{0}(i\omega+i\Omega,\bm{k}+\bm{q})D(i\Omega,\bm{q})
\notag \\
&= -g^{2}\frac{1}{\beta}\sum_{i\Omega}\int\frac{d^{2}q}{(2\pi)^{2}}
\frac{i(\omega+\Omega)+A(k_{x}+q_{x})^{2}\tau_{x}+v (k_{y}+q_{y})\tau_{y}}
{(\omega+\Omega)^{2}+A^{2}(k_{x}+q_{x})^{4}+v^{2}(k_{y}+q_{y})^{2}}
D(i\Omega,\bm{q}),
\end{align}
The self-energy has a constant part 
$
\Sigma(\omega=0, \bm{k}=\bm{0}) \approx - \alpha/(2\pi^2) v \Lambda \log \left( A \Lambda /v \right) \tau_x,
$
It does not have a divergence in the infrared limit, and hence this correction can be neglected in the low-energy analysis. 
Also it does not alter the discussion in the low-energy region. 
At $T=0$, the sum of Matsubara frequency turns to be integral of frequency along the imaginary axis. 
We introduce $\Sigma_\omega$, $\Sigma_{k_x}$, and $\Sigma_{k_y}$ to simplify the presentation as follows: 
\begin{align} 
\Sigma(\omega,\bm{k})
&= (-ig)^{2}\int \frac{d\Omega}{2\pi}\int\frac{d^{2}q}{(2\pi)^{2}}
G_{0}(i\omega + i\Omega,\bm{k}+\bm{q})D(i\Omega,\bm{q}) \notag \\
& = \Sigma_\omega \cdot i\omega - \Sigma_{k_x} \cdot Ak_x^2 \tau_x - \Sigma_{k_y} \cdot v k_y \tau_y.
\end{align}

\subsection{Evaulation of self-energy at $T=0$}
\subsubsection{$\Sigma_\omega$}

Under $k_x = k_y = 0$, we assume $|\Omega|,\, E_{\bm{q}} \gg |\omega|$ where $E_{\bm{q}}=\sqrt{A^{2}q_{x}^{4}+v^{2}q_{y}^{2}}$,
and expand $G_{0}(\omega+\Omega,\bm{q})$ with respect to $\omega$ up to the linear order.
Then we obtain
\begin{equation}
\Sigma_\omega =
g^2 \int \frac{d\Omega}{2\pi} \int \frac{d^2 q}{(2\pi)^2} \frac{\Omega^2 - E_{\bm{q}}^2}{(\Omega^2 + E_{\bm{q}}^2)^2} D(\Omega, \bm{q}).
\end{equation}
If $D(\Omega, \bm{q})$ is independent of $\Omega$, e.g., if we adopt $D_0(\bm{q})$, the integral over $\Omega$ yields zero because the integrand has no first-order pole.
We introduce dimensionless variables $x$ and $y$ as
\begin{equation}
x = \frac{\sqrt{A} q_x}{\sqrt{|\Omega|}}, \quad y = \frac{vq_y}{|\Omega|}
\end{equation}
to rewrite the integral. Then we have
\begin{align}\label{eqn:Sigma_w}
\Sigma_\omega
&= \frac{\alpha}{(2\pi)^3} \int \frac{d\Omega}{|\Omega|} \int dx dy \frac{1 - x^4 -y^2}{(1 + x^4 + y^2)^2} d(\Omega, x, y) 
\equiv \Sigma^{(0)} + \Sigma^{(1)}_\omega,
\end{align}
where $\Sigma^{(0)}$ and $\Sigma^{(1)}_\omega$ are defined by 
\begin{align}
\Sigma^{(0)} = \frac{\alpha}{(2\pi)^3} \int \frac{d\Omega}{|\Omega|} \int dx dy \frac{1}{(1 + x^4 + y^2)^2} d(\Omega, x, y), \notag \\
\Sigma^{(1)}_\omega = \frac{\alpha}{(2\pi)^3} \int \frac{d\Omega}{|\Omega|} \int dx dy \frac{- x^4 -y^2}{(1 + x^4 + y^2)^2} d(\Omega, x, y), 
\end{align}
and the function $d(\Omega, x, y)$ is given by
\begin{align}
d^{-1} (\Omega, x, y) 
= 2\sqrt{x^2 + \frac{|\Omega| Ay^2}{v^2}} 
+ \alpha_N \left[
\frac{d_x x^2}{(1 + cx^4 + y^2)^{1/4}}
+\frac{d_y y^2}{(1 + cx^4 + y^2)^{3/4}}
\right].
\end{align}
Note that $d(\Omega, x, y)$ is singular at $x=y=0$ even for $\alpha_N \gg 1$. 
Due to this singularity, $\Sigma^{(0)}$ and $\Sigma^{(1)}_\omega$ will be calculated separately.

To calculate $\Sigma^{(0)}$, we introduce the polar coordinate $x = r \cos\theta$, $y = r \sin\theta$.
In this case, the first term of $d^{-1} (\Omega, x, y)$ must be retained to avoid the singular behavior at $r = 0$.
Dominant contribution comes from small $r$ region.
For $\Omega$ integral, we introduce the upper energy cutoff $\Lambda$ and the lower cutoff $E$. 
Then we can approximate the integral as
\begin{align}
\Sigma^{(0)} 
& \approx  \frac{\alpha}{4\pi^3} \int_{E}^{\Lambda} \frac{d\Omega}{\Omega} \int_0^{2\pi} d\theta \int_0^1 dr d^{-1}(\Omega, x, y)\notag \\
& \approx  \frac{1}{4\pi^3} \frac{1}{N} \int_{E}^{\Lambda} \frac{d\Omega}{\Omega}
\int_0^{2\pi} \frac{d\theta}{d_x \cos^2\theta + d_y \sin^2\theta} 
\log \left( \alpha_N \frac{d_x \cos^2 \theta + d_y \sin^2 \theta}{2\sqrt{\cos^2 \theta + \frac{\Omega A}{v^2} \sin^2 \theta}} +1 \right)
\notag \\
&\approx  \left[ \frac{\sqrt{15}}{\pi^{3/2}} \frac{\log \alpha_N}{N} -\frac{f(\Lambda)}{N} \right]l, 
\end{align}
where $f(\Lambda)$ is a nonsingular function that depends on the cutoff and $l = \log (\Lambda / E)$ . 

Next we calculate $\Sigma^{(1)}_\omega$. Here $x^4 + y^2$ in the numerator compensate the singularity of $d(\Omega, x, y)$, and hence the large-$N$ limit can be safely taken before the integration.
Then we obtain
\begin{align}
\Sigma^{(1)}_\omega 
& \approx  - \frac{1}{4\pi^3 N} \int_{E}^{\Lambda} \frac{d\Omega}{\Omega} \int dx dy \frac{x^4 + y^2}{(1 + x^4 + y^2)^2} 
 \left[ \frac{d_x x^2}{(1 + c x^4 + y^2)^{1/4}} + \frac{d_y y^2}{(1 + c x^4 + y^2)^{3/4}} \right]^{-1} 
 = - \frac{0.5561}{N} l.
\end{align}
The integral with respect to $x$ and $y$ is obtained numerically.

\subsubsection{$\Sigma_{k_x}$}

We assume $|\Omega|,\, E_{\bm{q}} \gg Ak_x^2$ with $\omega = k_y = 0$ and expand $G_{0}(\Omega,k_{x}+q_{x},q_{y})$ up to $k_x^2$. 
Using the dimensionless parameters $x$ and $y$, we have
\begin{align}
\Sigma_{k_x} 
&= \frac{\alpha}{(2\pi)^3} \int \frac{d\Omega}{|\Omega|} \int dx dy 
 \frac{(1+y^2)^2 - 12 x^4 (1+y^2) + 3x^8}{(1 + x^4 + y^2)^3} d(\Omega, x, y) 
 \equiv \Sigma^{(0)} + \Sigma^{(1)}_{k_x}, 
\end{align}
with
\begin{align}
\Sigma^{(1)}_{k_x} = \frac{\alpha}{(2\pi)^3} \int \frac{d\Omega}{|\Omega|} \int dx dy 
 \frac{3x^8 -13x^4 -12x^4 y^2 +y^4 +y^2}{(1 + x^4 + y^2)^3} d(\Omega, x, y). 
\end{align}
In calculating $\Sigma^{(1)}_{k_x}$, the large-$N$ limit can be safely taken similarly as in $\Sigma^{(1)}_\omega$, and we obtain
\begin{align}
\Sigma^{(1)}_{k_x} & \approx  \frac{1}{4\pi^3 N} \int_{E}^{\Lambda} \frac{d\Omega}{\Omega} \int dx dy 
\frac{3x^8 -13x^4 -12x^4 y^2 +y^4 +y^2}{(1 + x^4 + y^2)^3} 
\left[ \frac{d_x x^2}{(1 + c x^4 + y^2)^{1/4}} + \frac{d_y y^2}{(1 + c x^4 + y^2)^{3/4}} \right]^{-1}
=  - \frac{0.4521}{N} l.
\end{align}

\subsubsection{$\Sigma_{k_y}$}

Finally we consider the case where $|\Omega|,\, E_{\bm{q}} \gg |v k_y|$ with $\omega = k_x = 0$ and expand $G_{0}(\Omega, q_x, k_y+q_y)$ up to the linear order in $k_y$.
Then we obtain 
\begin{align}
\Sigma_{k_y}
& = \frac{\alpha}{(2\pi)^3} \int \frac{d\Omega}{|\Omega|} \int dx dy \frac{1 + x^4 -y^2}{(1 + x^4 + y^2)^2} d(\Omega, x, y) 
\equiv \Sigma^{(0)} + \Sigma^{(1)}_{k_y}, 
\end{align}
where the function $\Sigma^{(1)}_{k_y}$ is defined by 
\begin{align}
\Sigma^{(1)}_{k_y}
= \frac{\alpha}{(2\pi)^3} \int \frac{d\Omega}{|\Omega|} \int dx dy \frac{x^4 -y^2}{(1 + x^4 + y^2)^2} d(\Omega, x, y) ,
\end{align}
and calculated similarly as $\Sigma^{(1)}_\omega$ and $\Sigma^{(1)}_{k_x}$ as
\begin{align}
\Sigma^{(1)}_{k_y} & \approx \frac{1}{4\pi^3 N} \int_{E}^{\Lambda} \frac{d\Omega}{\Omega} \int dx dy \frac{x^4 - y^2}{(1 + x^4 + y^2)^2} 
\left[ \frac{d_x x^2}{(1 + c x^4 + y^2)^{1/4}} + \frac{d_y y^2}{(1 + c x^4 + y^2)^{3/4}} \right]^{-1} 
= - \frac{0.2157}{N} l.
\end{align}

\section{\label{sec:RGstrongcoupling} Renormalization group analysis}

Now let us perform a renormalization group analysis.
The effective action is given by
\begin{align}
S & = \int_{k}\psi^{\dagger}_{k} \big( -i\omega + Ak_{x}^{2}\tau_{x} + v k_{y}\tau_{y} - \Sigma(k) \big) \psi_{k}
+ \int_{k,k'} ig \big( 1 + \delta \Gamma (k, k') \big) \psi^{\dagger}_{k} \psi_{k'} \phi_{k-k'}
+ \frac{1}{2} \int_{k}D^{-1}(\bm{k})\phi^{\dagger}_{k}\phi_{k}
\notag \\
& = \int_{k} \psi^{\dagger}_{k} \big( -i\omega (1+\Sigma_\omega) + A (1+\Sigma_{k_x}) k_{x}^{2}\tau_{x} + v (1+\Sigma_{k_y}) k_{y}\tau_{y}\big)\psi_{k} \notag \\
& \quad + \int_{k,k'} ig \big( 1 + \delta \Gamma (k, k') \big) \psi^{\dagger}_{k}\psi_{k'}\phi_{k-k'}
+ \frac{1}{2}\int_{k} D^{-1}(\bm{k})\phi^{\dagger}_{k}\phi_{k}.
\end{align}
The action changes by rescaling as
\begin{align}
\label{eq:rescaling1}
S &= b^{-1-z-z_{2}} Z_{\psi}^{-2} \int_{\tilde{k}} \tilde{\psi}^{\dagger}_{\tilde{k}}
\left[ -ib^{-z} \tilde{\omega} (1+\Sigma_\omega) + Z_{A}^{-1} b^{-2} \tilde{A} (1+\Sigma_{k_x}) \tilde{k}_{x}^{2} \tau_{x}
+ Z_{v}^{-1} b^{-z_{2}} \tilde{v} (1+\Sigma_{k_y}) \tilde{k}_{y}\tau_{y} \right] \tilde{\psi}_{\tilde{k}}
\notag \\
& \quad
+ b^{-2-2z-2z_{2}} Z_{\psi}^{-2} Z_{\phi}^{-1} Z_{g}^{-1} \int_{\tilde{k},\tilde{k}'} i\tilde{g} \tilde{\psi}^{\dagger}_{\tilde{k}} \tilde{\psi}_{\tilde{k}'} \tilde{\phi}_{\tilde{k}-\tilde{k}'}
+ \frac{1}{2} b^{-2-z-z_{2}} Z_{\phi}^{-2} \int_{\tilde{k}} D^{-1}(\tilde{\bm{k}}) \tilde{\phi}^{\dagger}_{\tilde{k}} \tilde{\phi}_{\tilde{k}} ,
\end{align}
where we put tildes to represent rescaled quantities.
Here we rescale the frequency and momenta as
\begin{equation}
\tilde{\omega}=b^{z}\omega, \quad \tilde{k}_{x}=bk_{x}, \quad \tilde{k}_{y}=b^{z_{2}}k_{y},
\end{equation}
and the fields and the other parameters are rescaled as
\begin{equation}
\tilde{\psi}=Z_{\psi}\psi, \quad \tilde{\phi}=Z_{\phi}\phi, \quad \tilde{A}=Z_{A}A, \quad \tilde{v}=Z_{v}v, \quad\tilde{g}=Z_{g}g.
\end{equation}
We note that 
$D^{-1}$ has the same scaling dimension as $k_x$.
The effective action keeps its form by the rescaling to be
\begin{align}
\label{eq:rescaling2}
S & = \int_{\tilde{k}} \tilde{\psi}^{\dagger}_{\tilde{k}} \big ( -i\tilde{\omega} + \tilde{A} \tilde{k}_{x}^{2}\tau_{x} + \tilde{v} \tilde{k}_{y}\tau_{y} \big) \tilde{\psi}_{\tilde{k}}
+ \int_{\tilde{k},\tilde{k'}} i\tilde{g} \tilde{\psi}^{\dagger}_{\tilde{k}} \tilde{\psi}_{\tilde{k'}}\tilde{\phi}_{\tilde{k}-\tilde{k'}}
+\frac{1}{2}\int_{\tilde{k}} D^{-1}(\tilde{\bm{k}}) \tilde{\phi}^{\dagger}_{\tilde{k}} \tilde{\phi}_{\tilde{k}},
\end{align}
Comparing Eqs.~\eqref{eq:rescaling1} and \eqref{eq:rescaling2}, we obtain the following relations:
\begin{align}\label{eqn:largeNscalefactor}
Z_{\psi}^2 &= b^{-1-2z-z_{2}} (1+\Sigma_\omega), \notag \\
Z_{\phi}^2 &= b^{-2-z-z_{2}}, \notag \\
Z_{A} &= b^{-3-z-z_2} Z_\psi^{-2} (1 + \Sigma_{k_x}) \approx b^{z-2} (1 + \Sigma_{k_x} - \Sigma_\omega) \approx 1 + (z-2+\gamma_A) l, \notag \\
Z_{v} &= b^{-1-z-2z_2} Z_\psi^{-2} (1 + \Sigma_{k_y}) \approx b^{z-z_{2}} (1 + \Sigma_{k_y} - \Sigma_\omega) \approx 1 + (z-z_2+\gamma_v) l, \notag \\
Z_{g} &= b^{-2-2z-2z_{2}} Z_{\psi}^{-2} Z_{\phi}^{-1} \big( 1 + \delta \Gamma (0,0) \big) \approx b^{(z-z_{2})/2} \big(1 + \delta \Gamma (0,0) - \Sigma_\omega \big)
\approx 1 + \frac{z-z_2}{2} l ,
\end{align}
where we define $\gamma_A$ and $\gamma_v$ as
\begin{gather}
\gamma_A = \frac{d}{dl} (\Sigma_{k_x}-\Sigma_\omega) \Big|_{l=0} \approx \frac{0.1261}{N}, \quad
\gamma_v = \frac{d}{dl} (\Sigma_{k_y}-\Sigma_\omega) \Big|_{l=0} \approx \frac{0.3625}{N}.
\end{gather}

\section{Charge screening}

Let us consider the static screening effect by introducing a charged impurity.
The charge distribution induced by the impurity is given by
\begin{align}
\rho_{\text{ind}}(\bm{q})& =ZeD(\omega=0,\bm{q})N\Pi(\omega=0, \bm{q}) \notag \\
&= -Ze \alpha_N \frac{B_{x}|q_{x}|+B_{y}\sqrt{|q_{y}|}}{2\sqrt{q_{x}^{2}+q_{y}^{2}}+ \alpha_N \left( B_{x}|q_{x}| + B_{y}\sqrt{|q_{y}|} \right) }, 
\end{align}
where the static polarization is 
\begin{align}
\Pi(\omega=0, \bm{q})=
-\frac{\alpha}{16}|q_{x}|-\frac{\alpha}{6\sqrt{\pi}}\frac{\Gamma(5/4)}{\Gamma(3/4)}\sqrt{\frac{v}{A}}|q_{y}|^{1/2}
\equiv - \alpha \left ( B_{x}|q_{x}| - \alpha B_{y}\sqrt{|q_{y}|} \right). 
\end{align}
The real space distribution of the screening charge $\rho_{\text{ind}}(\bm{r})$
can be obtained after Fourier transformation:
\begin{align}
\rho_{\text{ind}}(\bm{r})=\int \frac{d^{2}q}{(2\pi)^{2}}e^{i\bm{q}\cdot\bm{r}}\rho_{\text{ind}}(\bm{q}).
\end{align}
We define partially integrated charge densities as
\begin{gather}
Q_{x}(x)=\int dy \rho_{\text{ind}}(\bm{r})= \int \frac{dq_x}{2\pi} e^{iq_x x} \rho_{\text{ind}}(q_x ,q_{y}=0),
\nonumber\\
Q_{y}(y)=\int dx \rho_{\text{ind}}(\bm{r})= \int \frac{dq_y}{2\pi} e^{iq_y y} \rho_{\text{ind}}(q_x = 0 ,q_y).  
\end{gather}

\subsection{$\alpha_N \ll 1$}

For $N\alpha \ll 1$, the induced charge in momentum space is given by 
\begin{gather}
\rho_{\text{ind}}(\bm{q})= 
-Ze N\alpha \frac{B_{x}|q_{x}|+B_{y}\sqrt{|q_{y}|}}{2\sqrt{q_{x}^{2}+q_{y}^{2}}}, 
\end{gather}
and hence we obtain the partially integrated charge density
\begin{gather}
Q_{x}(x) = -\frac{1}{2} Ze N\alpha B_{x}\delta(x),
\quad
Q_{y}(y) = -\frac{1}{\sqrt{8\pi}} Ze N\alpha B_{y}\frac{1}{\sqrt{|y|}}. 
\end{gather}
Thus the integrated charge along the quadratic dispersion direction $Q_{x}(x)$ is localized
at the impurity site whereas the integrated charge along the linear dispersion direction $Q_{y}(y)$
has a power law decay.

\begin{figure}
\centering
\includegraphics[width=7cm]{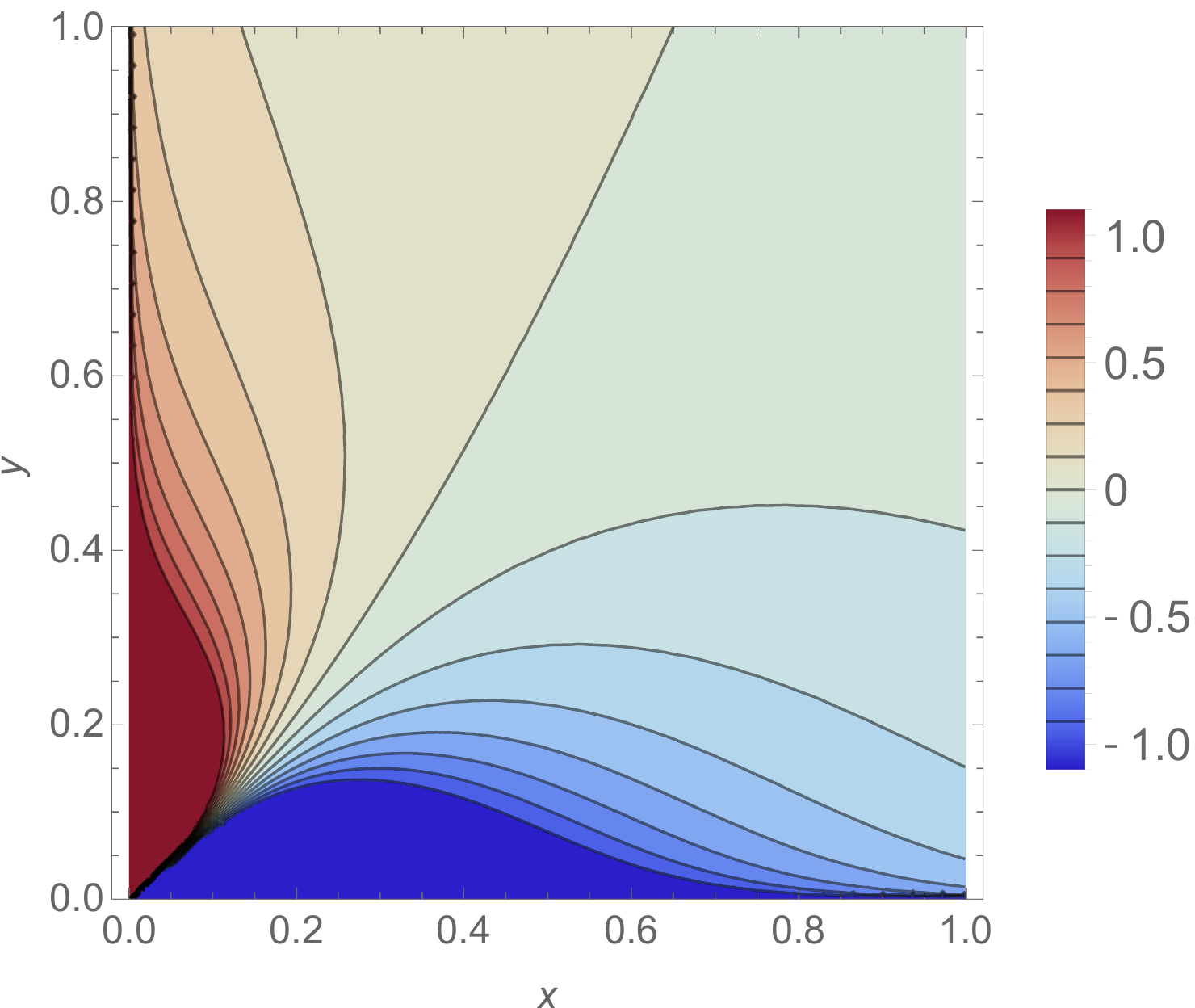}
\caption{
Induced charge distribution $\rho_{\text{ind}}(\bm{r})$. $x$ and $y$ are measured in unit of $(B_x/B_y)^2$ and we set $ZeN\alpha B_x^{-3} B_y^4/2 = 1$. 
}\label{fig:screening}
\end{figure}

The charge distribution in real space $\rho_{\text{ind}}(\bm{r})$ is obtained by the Fourier transform of $\rho_{\text{ind}}(\bm{q})$: 
\begin{equation}
\rho_{\text{ind}}(\bm{q}) = \frac{1}{2} Ze N\alpha [B_x h_1(\bm{r}) + B_y h_2(\bm{r})], 
\end{equation}
where $h_1(\bm{r})$ and $h_2(\bm{r})$ are defined as 
\begin{gather}
h_1(\bm{r}) = \frac{1}{\pi^2} \frac{1-\frac{x}{r} \sinh^{-1} \left(\frac{x}{|y|}\right)}{r^2}, \notag \\
h_2(\bm{r}) = -\frac{1}{\sqrt{2}\pi^{3/2} |x|^{3/2}} \frac{1}{1+\rho^2} \left\{
\text{Re} \left[ 2E\left( \frac{1+i\rho}{2} \right) -K \left( \frac{1+i\rho}{2} \right) \right]
- \rho \text{Im} \left[ K \left( \frac{1+i\rho}{2} \right) \right]
\right\}, 
\end{gather}
where $K(m)$ and $E(m)$ are the complete elliptic integrals of the first and second kind, respectively, with the parameter $m$, and we define $r=\sqrt{x^2 + y^2}$ and $\rho = |y/x|$. The charge distribution is shown in Fig.~\ref{fig:screening}. 

\subsection{$\alpha_N \gg 1$}

For $N\alpha \gg 1$, $\rho_{\text{ind}}(\bm{q}) = -Ze$ leads to $\rho_{\text{ind}}(\bm{q}) = -Ze \delta(\bm{r}) /(2\pi)^2$. 
The effect of the screening is large in this limit, and induced charge is localized on the impurity.

\section{Physical observables}

If we consider an observable $O$, we expect the scaling behavior 
\begin{equation}
O(\bm{k}, \omega, v, A) = z_O O(\bm{k}(l), \omega(l), v(l), A(l) ), 
\end{equation}
where $z_O$ is the scaling dimension of the operator $O$, and $\bm{k} (l) = (z_x(l) k_x, e^l k_y)$ and $\omega(l) = z_\omega (l) \omega$ are the running momenta and frequency, respectively. 
$z_x$ and $z_\omega$ obey
\begin{gather}
\frac{dz_x}{dl} = \frac{1}{2} z_x (1+\gamma_A-\gamma_v), \quad 
\frac{dz_\omega}{dl} = z_\omega (1-\gamma_v). 
\end{gather}
Defining $b = e^l$, we obtain
\begin{gather}
z_x = b^{(1+\gamma_A-\gamma_v)/2}, \quad z_\omega = b^{1-\gamma_v}.
\end{gather}

The scaling behavior discussed here is only valid for $\alpha_N \gg 1$. 
In the weak coupling case, the low-energy physics can be obtained via naive Hartree-Fock perturbation theory with weak coupling RG analysis. Here one expects logarithmic corrections, not corrections to power law behavior. 

\subsection{Compressibility}

In case of the particle density $O = n$, we have $z_{O=n} = b^{-1} z_x^{-1}$, such that the compressibility $\kappa = \partial n / \partial \mu$ follows $z_{O=\kappa} = b^{-1} z_x^{-1} z_\omega$. 
The scaling relation then implies
\begin{equation}
\kappa (T) = b^{-\gamma_v - (1+\gamma_A -\gamma_v)/2} \kappa (b^{1-\gamma_v} T), 
\end{equation}
which leads to 
\begin{equation}
\kappa(T) \propto T^{1/2+\phi}, 
\end{equation}
with 
\begin{equation}
\phi = \gamma_v + \frac{1}{2} \gamma_A = \frac{0.4255}{N}. 
\end{equation}

\subsection{Heat capacity}

The scaling of the free energy density is 
\begin{equation}
f(T, \mu) = b^{-1} z_x^{-1} z_\omega^{-1} f(z_\omega T, z_\omega \mu), 
\end{equation}
which reproduces the above scaling dimensions for the particle density via $n = \partial f / \partial \mu$. 
For the heat capacity $C = \partial f /\partial T$, its scaling dimension implies
\begin{equation}
C(T) = b^{-1} z_x^{-1} C(z_\omega T).
\end{equation}
Then we obtain 
\begin{equation}
C(T) \propto T^{3/2+\phi}.
\end{equation}

\subsection{Optical conductivity}

We first calculate first the bare optical conductivity. 
The current-current correlation function $\Pi_{\alpha\beta} (\bm{q}, \omega)$ $(\alpha, \beta = x, y)$
for the two direction is 
\begin{gather}
\Pi_{xx} (i\Omega) = -4N A^2 e^2 \int \frac{d\omega d^2k}{(2\pi)^3} k_x^2 \text{tr} [\tau_x G(\omega, \bm{k}) \tau_x G(\omega+\Omega, \bm{k})] , \\
\Pi_{yy} (i\Omega) = -N v^2 e^2 \int \frac{d\omega d^2k}{(2\pi)^3} \text{tr} [\tau_y G(\omega, \bm{k}) \tau_y G(\omega+\Omega, \bm{k})]. 
\end{gather}
Since we are only interested in the imaginary part of $\Pi_{\alpha\alpha}$ on the real axis, and thus we subtract the $\Omega = 0$ contribution. Then we obtain
\begin{gather}
\Pi_{xx} (i\Omega) - \Pi_{xx} (0) = -4N \Omega^2 \sqrt{A} e^2 \int \frac{d^2k}{(2\pi)^2} \frac{Ak_x^2 \cdot v^2k_y^2}{(A^2k_x^4 + v^2 k_y^2)^{3/2}} \frac{1}{\Omega^2+4A^2 k_x^4+4 v^2 k_y^2} , \\
\Pi_{yy} (i\Omega) - \Pi_{yy} (0) = -N \Omega^2 v^2 e^2 \int \frac{d^2k}{(2\pi)^2} \frac{A^2 k_x^4}{(A^2 k_x^4 + v^2 k_y^2)^{3/2}} \frac{1}{\Omega^2+4A^2k_x^4+4v^2k_y^2}. 
\end{gather}
With $x=\sqrt{A}k_x/\sqrt{|\Omega|}$ and $y = vk_y/|\Omega|$, it follows that 
\begin{gather}
\Pi_{xx} (i\Omega) - \Pi_{xx} (0) = -4 c_x N |\Omega|^{3/2} \frac{\sqrt{A}}{v} e^2 , \\
\Pi_{yy} (i\Omega) - \Pi_{yy} (0) = - c_y N |\Omega|^{1/2} \frac{v}{\sqrt{A}} e^2 , 
\end{gather}
where $c_x$ and $c_y$ are numerical constants, given by
\begin{gather}
c_x = \int_{-\infty}^\infty \frac{dxdy}{(2\pi)^2} \frac{x^2 y^2}{(x^4 + y^2)^{3/2}} \frac{1}{1+4x^4+4y^2}, \quad
c_y = \int_{-\infty}^\infty \frac{dxdy}{(2\pi)^2} \frac{x^4}{(x^4 + y^2)^{3/2}} \frac{1}{1+4x^4+4y^2}.
\end{gather}

It holds for $-1 < \eta < 1$ that $\Pi_{\alpha \alpha} (i\Omega) = -B |\Omega|^{1-\eta}$ leads to $\text{Im} \Pi_{\alpha \alpha} (\omega+i0^+) = B \cos (\pi \eta /2 ) \omega |\omega|^{-\eta}$. 
Thus, it follows for the real part of the optical conductivity 
\begin{equation}
\sigma_{\alpha\alpha} (\omega) = \frac{\text{Im} \Pi_{\alpha\alpha} (\omega+i0^+)}{\omega}
\end{equation}
that 
\begin{gather}
\sigma_{xx} (\omega) = N \frac{e^2}{\hbar} c'_x \left( \frac{\omega}{\omega_0} \right)^{1/2}, \quad
\sigma_{yy} (\omega) = N \frac{e^2}{\hbar} c'_y \left( \frac{\omega_0}{\omega} \right)^{1/2},
\end{gather}
with $\omega_0 = v^2/A$ and the constants $c'_x = 4c_x/\sqrt{2} \approx 0.05$, $c'_y = c_x/\sqrt{2} \approx 0.05$.

Using the gauge invariance, one finds that the scaling dimensions of the optical conductivity are different for the two directions:
\begin{gather}
\sigma_{xx} (\omega) = b^{-1} z_x \sigma_{xx} (z_\omega \omega), \quad
\sigma_{yy} (\omega) = b z_x^{-1} \sigma_{yy} (z_\omega \omega). 
\end{gather}
Then it follows that 
\begin{gather}
\sigma_{xx} (\omega) \propto N \frac{e^2}{\hbar} \left( \frac{\omega}{\omega_0} \right)^{1/2+\phi_\sigma}, \quad
\sigma_{yy} (\omega) \propto N \frac{e^2}{\hbar} \left( \frac{\omega}{\omega_0} \right)^{-1/2-\phi_\sigma},
\end{gather}
with $\phi_\sigma = \gamma_v -\gamma_A/2 \approx 0.299/N$. 
The anisotropy of the optical conductivity is amplified by strong interactions.

\subsection{Diamagnetic susceptibility}

We consider the diamagnetic response to a field perpendicular to the plane. 
We use for the diamagnetic susceptibility
\begin{equation}
\chi_D = \lim_{\bm{q}\to 0} \frac{1}{q_x q_y} \Pi_{xy} (\bm{q}, \omega =0),
\end{equation}
with the current-current correlation function $\Pi_{\alpha\beta} (\bm{q}, \omega)$ $(\alpha, \beta = x, y)$. 
The scaling dimension of $\Pi_{xy}$ is $z_\omega^{-1}$. This follows from the following logic:
Using the Kubo formula and our results for the conductivity it holds that $\Pi_{xx}$ has a scaling dimension $b^{-1} z_x z_\omega^{-1}$ and $\Pi_{yy}$ has $b z_x^{-1} z_\omega^{-1}$. 
Thus, a mixed term $\Pi_{xy}$ must have the geometric mean as scaling dimension, i.e., $z_\omega^{-1}$. 
It leads to the scaling relation 
\begin{equation}
\chi_D (T) = b z_x z_\omega^{-1} \chi_D (z_\omega T), 
\end{equation}
which yields 
\begin{equation}
\chi_D \propto T^{-1/2-\phi},
\end{equation}
with $\phi = \gamma_v + \gamma_A/2 = 0.4255/N$.


\begin{thebibliography}{99}
\bibitem{extra} Such an anisotropic momentum dependence of the polarization function
can induce a highly anisotropic distribution of screening charges
once a charged impurity is introduced to the system (see SM for details).
\end{thebibliography}
\end{document}